# Insights Informed Generative AI for Design: Incorporating Real-world Data for Text-to-Image Output

Richa Gupta[1], Alexander Htet Kyaw[1]

[1] Massachusetts Institute of Technology

**Abstract.** Generative AI, specifically text-to-image models, have revolutionized interior architectural design by enabling the rapid translation of conceptual ideas into visual representations from simple text prompts. While generative AI can produce visually appealing images they often lack actionable data for designers In this work, we propose a novel pipeline that integrates DALL-E 3 with a materials dataset to enrich AI-generated designs with sustainability metrics and material usage insights. After the model generates an interior design image, a post-processing module identifies the top ten materials present and pairs them with carbon dioxide equivalent ($CO_2e$) values from a general materials dictionary. This approach allows designers to immediately evaluate environmental impacts and refine prompts accordingly. We evaluate the system through three user tests: (1) no mention of sustainability to the user prior to the prompting process with generative AI, (2) sustainability goals communicated to the user before prompting, and (3) sustainability goals communicated along with quantitative $CO_2e$ data included in the generative AI outputs. Our qualitative and quantitative analyses reveal that the introduction of sustainability metrics in the third test leads to more informed design decisions, however, it can also trigger decision fatigue and lower overall satisfaction. Nevertheless, the majority% of participants reported incorporating sustainability principles into their workflows in the third test, underscoring the potential of integrated metrics to guide more ecologically responsible practices. Our findings showcase the importance of balancing design freedom with practical constraints, offering a clear path toward holistic, data-driven solutions in AI-assisted architectural design.

## 1    Introduction

Advances in Generative Artificial Intelligence (AI)  models are becoming increasingly adopted in architectural design.. Various socio-cultural trends have influenced architecture, either manifesting in building design or shaping the tools architects use. Recent advances in Generative AI tools aim to generate outputs for a variety of design tasks such as performance-based analysis, form-finding, spatial programming, multi-objective optimization, restoration, and design tool development has emerged [1] (see Fig. 1).

While these tools are being explored across many domains, this paper focuses specifically on their application in interior architectural design. The use of generative AI models in the field is reshaping traditional workflows and enhancing both efficiency and creative freedom. Specifically, text-to-image generative AI models, such as DALL-E 3 and Midjourney, are now widely used for visualization and rendering in interior architectural design [4,8]. Despite these advancements, current text-to-image models in the interior design domain are limited to a visual representation, without metrics, insights or data that would enable decision-making in the early phase of the design. The lack of real world data in generative AI outputs such as sustainability, cost, or material properties makes them difficult to apply in practical design contexts. Additionally, the absence of meaningful data restricts the ability to refine designs and iterate prompts effectively with generative AI.

This paper proposes a novel approach that shifts the role of generative AI from purely visual output generation to a decision-support system enriched with actionable insights. Such insights may include factors like material cost, $CO_2e$ emissions, porosity, VOCs, fire ratings, regional availability, and LEED credits [14, 15, 16]. However, for the purpose of this study, the scope is narrowed specifically to material-related insights, focusing on $CO_2e$ metrics associated with generative AI outputs.

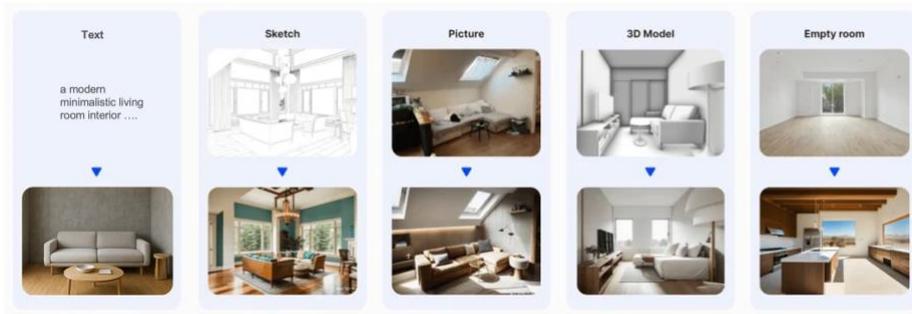

**Fig. 1.** Speculative Usage of Generative AI Tools for Interior Design Workflow [13]

Research demonstrates that decisions made during early design stages have a profound impact on environmental outcomes, yet current evaluation methods often neglect sustainability metrics [7]. To test the hypothesis that integrating material insights into generative design outputs can support more informed decision-making, this study limits the insights to the $CO_2e$ values of identified materials. Since Generative AI workflows are majorly used for the early design phase in the industry, sustainability metrics proves to be a firm choice for proving the hypotheses. The approach allows designers to iterate and refine their prompts in the Generative AI design workflow with sustainability metrics from the earliest stages. We further investigate the impact of these sustainability metrics on human decision-making, examining whether including such metrics leads to more satisfactory and responsible designs. By merging materials, sustainability, and generative models into one streamlined workflow, the ultimate aim is to empower architects with informed Generative AI Outputs.

## 2     Related Works

### 2.1     Gen AI in Architecture Design

Generative AI (Gen AI) has significantly expanded the possibilities for computer-assisted design in architecture, particularly through text-to-image modeling. Tools such as DALL-E enable designers to rapidly transform conceptual ideas into speculative visual representations by using natural language prompts. This bridging the often-lengthy gap between ideation and visualization [8]. This rapid transition from concept to representation allows designers to create compelling visuals and explore new ideas without relying on specialized rendering skills. Moreover, as AI models become more specialized, studies like those by Kahraman et al. [3] indicate that integrating Gen AI into academic curricula and professional workflows can prepare future architects to meet evolving industry demands. By aligning diffusion models according to stylistic or functional criteria, Gen AI can yield outputs that cater to specific client needs or thematic requirements, creating a new paradigm of human-AI co-creation in architectural design.

### 2.2     Design Cognition and Feedback-Based Insights

While generative models offer rapid visual outputs, effective architectural design is inherently iterative and relies heavily on continuous feedback. Schön's notion of "reflection-in-action" underscores the importance of an ongoing conversation between the designer and the evolving design artifact [7]. In practical terms, this means that architects engage in repeated cycles of critique, evaluation, and adjustment to refine the design toward a coherent goal. Rittel and Webber's concept of "wicked problems" further emphasizes the complexity of design tasks, which often involve open-ended challenges with no single "correct" solution [6]. As a result, the design process benefits from structured feedback loops. In this paper, we argue that such feedback loops can be integrated into the generative AI design process. Building on early design cognition research, we present a workflow in which AI-generated outputs are accompanied by real-world data as feedback mechanisms. This enables designers to reflect and iteratively refine their prompts based on data-driven feedback, allowing them to steer the outputs toward their desired goals. Such feedback-driven interactions also allow users to build trust in AI tools by progressively verifying and aligning outputs with real-world data and design objectives.

### 2.3     Importance of Information in Early Design Phase for Sustainability

Sustainability considerations have become increasingly central in contemporary architecture, as early-stage decisions can profoundly influence a project's long-term environmental impact [2]. Despite this recognition, they are missing in early phases of design. Studies reveal that when sustainability is treated as an afterthought, crucial opportunities for optimization may be lost, especially since later modifications to a design are more resource-intensive and less effective [5]. By embedding quantitative performance indicators directly into the early design phase, exemplified by tools like Grasshopper and EnergyPlus, architects can iteratively refine forms, materials, and layouts to align with measurable sustainability benchmarks.[19] While this has been done in conventional 3D modeling, the idea of embedding quantified performance integration into an AI-driven design process has not been explored before. Current explorations with generative AI have primarily focused on generating context-aware visuals and fine-tuning models to produce outputs in a specific style, with an emphasis on appearance and aesthetics instead of connection to real world data or sustainability. [20]

As Gen AI continues to evolve, this paper presents a growing expectation that such tools will not only propose novel design forms but also provide measurable environmental insights that guide architects toward responsible solutions. Achieving this synergy where generative outputs are backed by relevant sustainability metrics represents a critical frontier in advancing both the creative and practical dimensions of architectural design.

## 3  Method

We present a workflow that enables interior architectural designers to use generative AI to produce images accompanied with $CO_2e$ data linked directly to the generated outputs. The aim is to examine whether these insights influence users to prompt the AI toward more sustainable design directions or to choose more different materials.

Interior architectural design is selected as the primary domain for two reasons. First, interior spaces often involve intricate details including surface finishes, furniture, and fixtures that present an ideal application for text-to-image generative models. Second, decisions about materials and finishes in interior design have significant implications for sustainability. By concentrating on interior contexts, we can assess the efficacy of embedding $CO_2e$ metrics in a setting where aesthetic, functional, and environmental considerations intersect.

The method is divided into two parts: system implementation and user experiments.

### 3.1 System Implementation

Generative Design Intelligence: Extracting Design Specific Sustainability Insights and Material Metrics from Generative Artificial Intelligence output      5

There are multiple generative AI models that can be used in the design process. In this paper, we focus on a text-to-image generative model, specifically DALL·E 3, which is one of the latest models capable of producing high-fidelity images from user text prompts. A core objective of this work is to integrate sustainable design goals into the early phases of the design process when using generative AI. We use carbon dioxide equivalent ($CO_2e$) as a key metric to emphasize the environmental impact of material choices. By providing this metric along with the generative AI image output, the system enables designers to make informed prompt adjustments based on the data as feedback.

We propose a post-generative framework designed to augment image outputs from generative AI with interpretable, material-specific insights such as $CO_2e$. The pipeline involves five key components: a generative image model, a vision-language material descriptor, a material mapping agent, a database connector, and an insights interface. Together, the system provides a feedback mechanism that grounds speculative AI-generated images with real world material data.

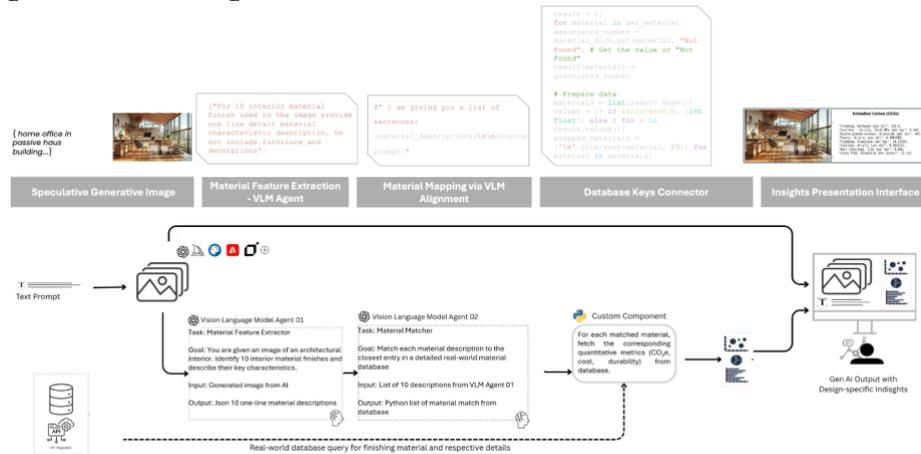

**Fig. 2.** Technical Implementation of the approach has five modular parts: a generative image model, a vision-language material descriptor, a material mapping agent, a database connector, and an insights presentation interface

**AI Generated Image.** The workflow begins with the user providing a structured text prompt that can include descriptions such as such as room types, layout, dimensions, lighting conditions, aesthetic, style, color palette, or material choices. This prompt is processed through a state-of-the-art text-to-image diffusion model (e.g., DALL·E 3), which generates a high-resolution image depicting an interior spatial proposal. This image serves as the foundational input for all subsequent reasoning tasks.

**Vision Language Material Feature Extraction .** The generated image is encoded into base64 format and analyzed by a vision-language model (VLM) tasked with identifying and describing ten interior material finishes. In this implementation we use GPT 4o by OpenAI as the vision language model. The model is prompted to exclude furniture and decorative elements, focusing exclusively on architectural and construction-related materials. For each material identified by the vision-language model, the model generates a one-line textual description detailing its visual characteristics and functional implications.

**Material Mapping via VLM Alignment.** These material descriptions are then passed to a second VLM agent that performs few-shot similarity matching between each description and a curated database of real-world construction materials (in this approach- Material2050 Database). The database includes taxonomically organized entries with consistent nomenclature and units (e.g., "Cladding, Limestone (per kg)"). The model selects the most semantically aligned material from the database.

```
{'id': 9,
 'material_name': 'Decking, Hardwood (per m3)',
 'group_elements_nrm_1': ['3 - Internal finishes', '8 - External works'],
 'elements_nrm_1': ['3.2 - Floor finishes', '8.X - External works'],
 'product_type': 'Decking',
 'product_type_family': ['Exterior finishes'],
 'uniclass_systems': ['(Ss_30_20) Flooring and decking systems',
  '(Ss_30_20_30_25) Decking systems'],
 'uniclass_products': ['(Pr_25_71_97) Wood-based boards'],
 'material_type': 'Hardwood',
 'material_type_family': 'Wood',
 'data_source': '2050 Materials Research (September 2022) - ARCHIVED',
 'functional_unit_quantity': '1',
 'functional_unit_unit': 'm3',
 'total_biogenic_co2e': -762.8,
 'carbon_a1a3': 226.8,
 'carbon_a5': 47.6,
 'carbon_c1c4': 1772.84,
 'freshwater_use_a1a3': 462.0,
 'reuse_potential': 50.0,
 'odp': '3.63e-10',
 'density': 570.5,
 'created': '2022-11-10T21:47:30.392000Z',
 'updated': '2024-08-21T20:15:38.889339Z'},
```

**Fig. 3.** Example of a datapoint within the 2050 Materials general materials dataset

**Database Connector.** Once a material is identified by the VLM agent, it is passed to a database connector that queries the Material2050 API to fetch real-world data, such as $CO_2e$ metrics. These numerical values are then presented alongside the AI-generated image, providing users with actionable insight into the environmental impact of the materials rendered in the AI generated image.

**Insights Interface.** Finally, the AI-generated image, along with its post-processed information (identified materials and corresponding $CO_2e$ scores), is presented to the user through a Google Colab interface. This enables users to evaluate the design and refine their prompts based not only on visual feedback, but also on sustainability metrics grounded in real-world data.

### 3.1 Experiments and User Studies

**Three Designed Tests.** To systematically evaluate the influence of material $CO_2e$ insights in interior design during the early phase while using Generative AI models, three tests for user study were designed involving professional interior designers and design students. For each test, the user goes through five design iteration attempts. After each AI-generated image is created, the user can view the output and adjust their prompt based on the feedback, progressively refining their design for five times.

*Test 1 - No Sustainability Consideration.* Participants were asked to generate interior design images using text-to-image model with no mention of sustainability or material metrics. This test functions as a control, establishing baseline design preferences without environmental constraints.

*Test 2 - Sustainability Goal (No Metrics).* Here, participants were instructed to integrate sustainability into their design objectives, yet no $CO_2e$ values or material types were displayed after image generation. This phase isolates the effect of merely stating a sustainability goal, without any quantitative metrics.

*Test 3 - Sustainability Goal with $CO_2e$ Metrics.* In this test, participants again aimed to create sustainable interior designs. This time, however, they were provided with the top 10 identified materials - alongside each material's associated $CO_2e$ values. By comparing the outcomes of Tests 1 and 2 with Test 3, the goal is to understand how actionable metrics and real-world data influence the user's decision-making process as they adjust their prompt during each attempt.

Table 1. Summary of designed three experimental tests.

| Test | Objective | Sustainability Goal | Metrics Provided |
|---|---|---|---|
| 1 | No sustainability considerations | No | None |
| 2 | Sustainability as a design goal | Yes | None |

8    Richa Gupta and Alexander Htet Kyaw

| Test | Objective | Sustainability Goal | Metrics Provided |
|---|---|---|---|
| 3 | Sustainability with goal & metrics | Yes | $CO_2e$ values, top 10 materials |

The study includes nine design professionals and students, where each participant completes all three tests. For each test, participants had five attempts to iterate their prompts to further refine the design After all five iterations were generated, participants completed a short qualitative survey capturing their impressions on design quality, ease of use, and in test 03- the perceived utility of $CO_2e$ metrics.

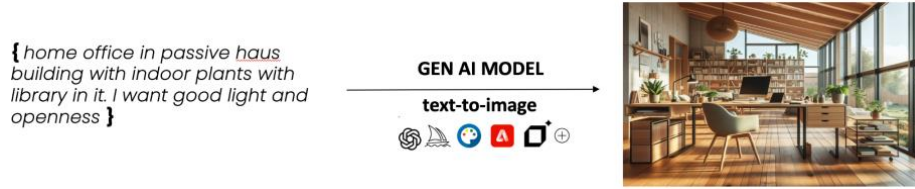

**Fig. 4.** Existing Text-to-Image (T-2-I) Pipeline. This existing pipeline was utilized in Test 01 and Test 02 i.e. no material $CO_2e$ insights.

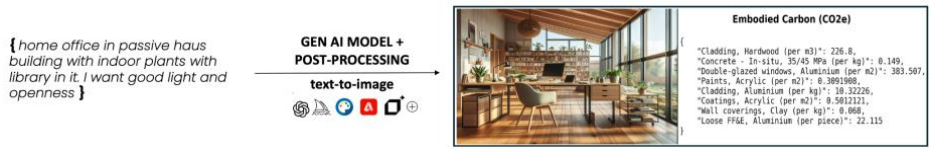

**Fig. 5.** Approach of this paper Text-to-Image + Insights (T-2-II) Pipeline. This was utilized in Test 03 i.e. material $CO_2e$ insights alongside Generative AI Output.

**User Study interface.** To facilitate structured, replicable evaluation of generative design behavior, three Google Colab notebooks were developed for each test as the primary interface for the user study. The notebook is designed to guide participants through a five-step iterative image generation process, using natural language prompts to progressively refine an interior design aligned with their stated goals.

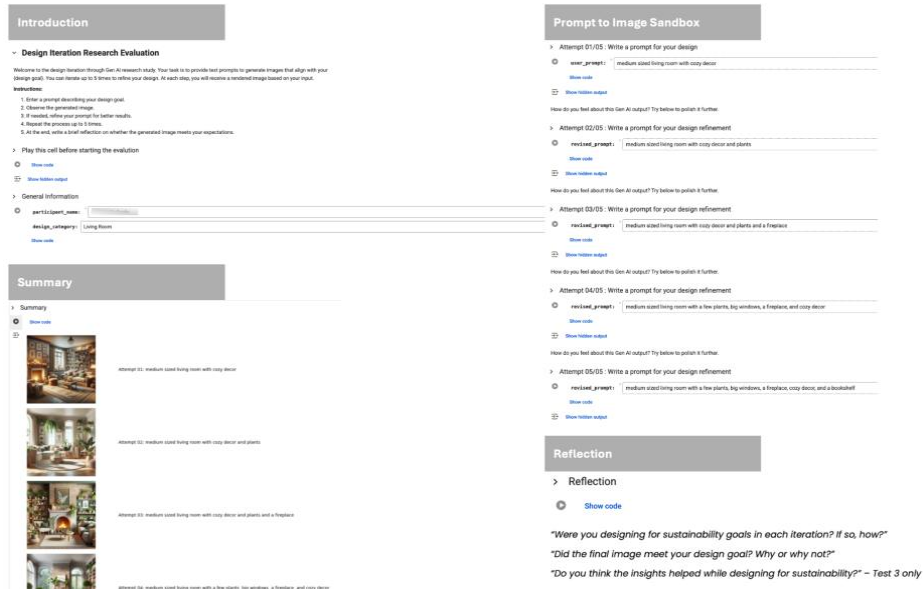

**Fig. 6.** Notebook for User Study with four primary components: Introduction, Data Collection, Summary and Reflection.

The interface comprised four sections:

*Introduction:* Participants were welcomed to the study and asked to input their design background and consent for the study. They were provided a brief introduction that they will be instructed to generate up to five design iterations and reflect on their evolving design logic in three different generative AI workflows.

*Prompt to Image Sandbox:* This is the primary experiment section where the users enter their text prompts and receive image outputs generated by the AI model. In the case of Test 3, the output also includes $CO_2e$ metrics for the identified materials. They then have the opportunity to revise their prompt based on the system output. This process is repeated for a total of five iterations. After each iteration, participants also provide a brief reflection on how well the output aligns with their original design intent.

*Summary:* The notebook automatically compiled the prompt-image pairs from all five attempts, providing a visual comparison and assessment for the user.

*Reflection:* At the conclusion, participants responded to open-ended questions about sustainability alignment, final design satisfaction, and (when applicable) whether AI-generated material insights supported their design process (Test 03 only).

**Evaluation.** This study employed a qualitative evaluation framework to examine how designers engaged with generative tools when presented with Insights. Participants' evolving prompt strategies and written reflections were analyzed to assess how material $CO_2e$ insights influenced their design intent and perceived success. Reflections were manually coded using a three-point scale (Yes = 1, Somewhat = 0.5, No = 0) to quantify design satisfaction, sustainability consideration, and perceived helpfulness of insights, providing a consistent basis for comparing user feedback across conditions.

While the present paper emphasizes qualitative trends, a more detailed quantitative evaluation using prompt similarity scores, lexical analysis, and standardized indices such as the Creativity Support Index (CSI)[18] and NASA-TLX [17] is planned for future work.

## 4      Results

The evaluation revealed different user responses across the three experimental tests, offering insight into how sustainability prompts and embedded material metrics affect user's prompts.

In **Test 1**, where participants were not given any sustainability goals or metrics, approximately 72% reported being satisfied with their final design outputs. Additionally, most users reported that they did not include sustainability as a design objective in their prompts. Qualitative responses focused primarily on visual and thematic alignment - describing outputs as "cool" or "aesthetically aligned" - highlighting that in the absence of design constraints or guidance, users relied on surface-level aesthetics without consideration for sustainability. This test served as a baseline, revealing a passive engagement with Generative AI when unassisted by either goal framing or data.

**Test 2** introduced sustainability as a design goal but no specific material metrics. This small intervention yielded a substantial behavioral shift: ~83% of participants reported incorporating sustainability considerations and satisfaction levels rose slightly to ~77%. Participants engaged with the task by integrating symbolic or intuitive strategies such as emphasizing natural light, greenery, or minimalist decor to approximate sustainable design. Reflections frequently referenced a "feeling of greenness" or "eco-aesthetic" rather than specific material trade-offs, suggesting that while goal-oriented framing increases awareness, it does not necessarily lead to grounded or materially actionable decisions.

In **Test 3**, where participants were provided both a sustainability goal and real-time insights in the form of $CO_2e$ values and material breakdowns, sustainability was factored into 100% of design decisions. Prompts became significantly more specific, with users incorporating language such as "rammed earth," "mass timber," and "low-carbon substitutes." Despite this increased alignment with the environmental goal, satisfaction dropped to ~50%. Participants reported experiencing higher cognitive load, decision fatigue, and difficulty balancing their aesthetic goals with the newly introduced constraints. While 66% of users found the material insights helpful describing them as "a good way to summarize impact" while others expressed concerns about the lack of a comparative baseline, insufficient context for the metrics, or identified limitations in the AI's ability to interpret nuanced material intent. Comments such as "AI not understanding material intention" and "forced into carbon logic" reflected the complexity of integrating quantitative feedback into creative workflows.

**Table 2.** Summary of Participant Responses Across Three Experimental Tests

| Test | Sustainability Considered (%) | Satisfaction (%) | Insights Found Useful (%) | Qualitative Behavior |
|---|---|---|---|---|
| **Test 1** (No sustainability framing) | ~0-10% | ~72% | N/A | Focused on aesthetics only; no mention of sustainability unless prompted. Responses described outputs as "cool" or "looks right." |
| **Test 2** (Sustainability goal only) | ~83% | ~77% | N/A | Incorporated symbolic sustainability (e.g., natural light, minimalism); referenced "feeling green" or comfort-based aesthetics. |
| **Test 3** (Sustainability goal + $CO_2e$ metrics) | ~100% | ~50% | ~66% | Prompt language became materially specific (e.g., "rammed earth," "mass timber"); some users expressed frustration over trade-offs and data interpretation. |

## 5  Discussion

Collectively, these findings suggest that while sustainability prompts alone can influence design intent, the inclusion of interpretable and actionable metrics leads to more deliberate prompt iterations and materially grounded design decisions. However, this benefit comes at the cost of increased cognitive burden and reduced design satisfaction when users are required to navigate complex trade-offs. This underscores the importance of calibrating the accuracy, interpretability, and framing of generative feedback in design tools. For generative systems to meaningfully support responsible interior decision-making, they must not only provide data but do so in a way that preserves creative agency and supports informed, user-centered choices.

### 5.1  Challenges

A key challenge identified during evaluation relates to the measurement unit inconsistencies in the general materials dictionary. The dictionary included materials measured in variety of units such as cubic meters, square meters, and kilograms, with each material's $CO_2e$ emission calculated correspondingly. This lack of standardization made it difficult for users to interpret the emission values and assess whether a given $CO_2e$ score was high or low in relative terms. A seemingly small value might have resulted from a smaller unit (e.g., kilograms), while a higher value could have been tied to a larger unit (e.g., cubic meters). Standardizing all materials into a single unit is not straightforward either, given that $CO_2e$ emissions often do not scale linearly due to fixed operational cost or manufacturing overheads (e.g., start-up energy for production machinery).

A second issue stems from dictionary granularity. In some cases the dictionary was too broad, failing to capture specific variations of a material. In other cases, it was too narrow, identifying closely related materials as the same. For instance, the system might detect "hardwood cladding" and "wood fiber floorboards," which have minimal aesthetic differences but vastly different functions and $CO_2e$ impacts. Cross referencing between the material dictionary, the prompt, and the image generation process could mitigate these inaccuracies. However, current text-to-image models like DALL-E 3 do not output both images and text-based descriptions of the precise materials used, limiting the potential for fully automated, high-fidelity material matching.

### 5.2  Future Works

**Data Objectives and Metrics**. Integrating a dataset with consistent units for materials would make emission values more interpretable. However, sustainability metrics are not the only considerations made while designing Interiors using generative AI tools. These metrics can be cost, durability, procurement, etc as discussed in the Introduction section. This progression will move the system from a speculative aid to a robust decision-support tool within generative AI design workflows.

**User Study.** A detailed quantitative analysis can be conducted using prompt similarity metrics, lexical variation, and standardized indices such as the Creativity Support Index (CSI)[18] and NASA-TLX [17] to assess user effort, cognitive load, and creative agency. These metrics will allow for a more rigorous comparison of design behavior across conditions.

**Real-World Applicability**. Bridging the gap between AI-generated designs and practical implementation remains a challenge. Future tools might link design outputs to specific materials and their carbon footprints, allowing easy translation from visual concepts to real-world projects. Additionally, extending the integration of insights with generative AI outputs beyond 2D images to 3D design software could significantly broaden the impact of these tools in practice.

**Responsible AI Parallels.** This approach offers strong alignment with Responsible AI principles by providing a lightweight, post-generative framework that has the potential to enhance transparency, user agency, and adaptability in early-stage design. Unlike end-to-end generative models, which may overlook regional material availability, user preferences, or cultural nuances, this method allows designers to refine generative AI outputs through interpretable, data-driven insights. Through post processing generative AI outputs, our method can provide meaningful sustainability metrics without the need to retrain foundational models. Future work could extend this framework to include additional criteria such as cost, durability, or ethical sourcing, further supporting a more holistic and accessible path toward responsible, user-aligned decision-making in AI-assisted design.

Additionally, this approach addresses an emerging ethical tension in generative design workflows: the creation of false expectations between designers and clients. AI-generated images frequently convey idealized visions that don't include material constraints, sustainability and construction feasibility, resulting in misaligned expectations. By embedding $CO_2e$ metrics and material references into AI generated outputs, this framework introduces a feedback mechanism that anchors design concepts in actionable, verifiable data. This not only empowers designers to communicate trade-offs more effectively but also fosters shared understanding, helping clients make informed choices that balance aesthetic aspiration with environmental responsibility supporting more trustworthy and grounded design collaborations.

## 6  Conclusion

In this paper, we introduced an AI-driven design pipeline that integrates sustainability metrics and material type information within a text-to-image generative AI model for interior architectural design. By leveraging DALL-E 3 and a general materials dataset, we provided users with $CO_2e$ metrics for up to ten materials in their generated designs. The qualitative results illustrate how this additional information significantly shapes user behavior, encouraging more iterative prompt refinements and deeper consideration of environmental impact.

Our findings demonstrate the potential and importance of including actionable sustainability data into generative AI design workflows. By offering users with $CO_2e$ metrics for AI generated designs, we promote a more informed, user-centric design approach. Moving forward, extending the system to include additional metrics (e.g., cost, resource availability, and life-cycle impact) promises to further enhance its relevance to architects, designers, and stakeholders seeking to balance creativity with real-world constraints in generative AI design workflows

**Acknowledgments.** This study was partly developed during Coursework at MIT named AI, Decision Making and Society with a team - Vin Baker and Richard Gu alongside authors. Grateful for advisors Dr. Terry Knight and Dr. Takehiko Nagakura.

## References

1. Bölek, B., Tutal, O., Özbaşaran, H.: A systematic review on artificial intelligence applications in architecture. *Journal of AI in Architecture* **2**(5), 99-110 (2023)
2. Han, J., Jiang, P., & Childs, P. (2021). Sustainability metrics in early-stage conceptual design: Impacts on environmental outcomes. *Design Studies, 72*(3), 105-122. https://doi.org/ 10.1016/j.destud.2021.01.007
3. Kahraman, M. U., S¸ekerci, Y., Develier, M., & Koyuncu, F. (2024). Integrating Artificial In- telligence in Interior Design Education: Concept Development. *Journal of Computational De- sign, 5*(1). https://doi.org/10.53710/jcode.1418783
4. Kimbell, L. (2012). Rethinking design thinking: Part II. *Design Culture, 4*(2), 129-148.
5. Meng, Z., Zhang, X., & Wei, Y. (2024). Design for a sustainable future: A review of sustainability-driven practices in the design community. *International Journal of Design, 18*(4), 45-61.
6. Rittel, H. W. J., & Webber, M. M. (1973). Dilemmas in a general theory of planning. *Policy Sciences, 4*(2), 155-169. https://doi.org/10.1007/BF01405730
7. Scho¨n, D. A. (1983). The Reflective Practitioner: How Professionals Think in Action. Basic Books.
8. Taiwo, A., Hu, B., & Zhao, J. (n.d.). Bridging conceptualization and visualization in architec- ture using AI-based tools like DALL-E. *AI & Architecture Journal*.
9. Abrishami, S., Goulding, J., Rahimian, F., & Ganah, A. (2014). Generative design in BIM: Automated solutions for early design stages. *Automation in Construction, 39*, 15-29. https://doi.org/10.1016/j.autcon.2014.07.002
10. Ma, Y., Li, W., & Chen, X. (2021). The role of generative design in optimizing architectural workflows. *Computers in Architecture, 18*(6), 89-97.
11. Sherwin, C., & Bhamra, T. (1999). Innovation through sustainability: A review of revised design approaches. *Journal of Product Innovation Management, 16*(5), 379-395.
12. Chertow, M. R. (2000). Industrial symbiosis: Literature and taxonomy. *Annual Review of Energy and Environment, 25*, 313-337. https://doi.org/10.1146/annurev. energy.25.1.313
13. Studio Hinton: Using AI for Interior Design: Pros and Cons. https://www.studiohinton.com/design-matters/using-ai-for-interior-design-pros-and-cons/
14. Jung, C., Abdelaziz Mahmoud, N.S., Al Qassimi, N. & Elsamanoudy, G. (2023) 'Preliminary study on the emission dynamics of TVOC and formaldehyde in homes with eco-friendly materials: beyond green building', Buildings, 13(11), 2847. https://doi.org/10.3390/buildings13112847
15. Riggio, M., Dalle Vedove, A. & Piazza, M. (2024) 'Load-bearing furniture modules for fast deployable and reusable systems', Frontiers in Built Environment, 10, 1405500. https://doi.org/10.3389/fbuil.2024.1405500
16. Escobar, I., Orduna-Hospital, E., Aporta, J. & Sanchez-Cano, A. (2024) 'Efficient daylighting: the importance of glazing transmittance and room surface reflectance', Buildings, 14(10), 3108. https://doi.org/10.3390/buildings14103108

17. Hart, S. G. (2006). NASA-task load index (NASA-TLX); 20 years later. Proc. Hum. Factors Ergon. Soc. 50, 904-908. doi: 10.1177/154193120605000909
18. Cherry, E., and Latulipe, C. (2014). Quantifying the creativity support of digital tools through the creativity support index. ACM Trans. Comput. Hum. Interact. 21, 1-25. doi: 10.1145/2617588
19. Brown, N., & Mueller, C. T. (2016). Design for Structural and Energy Performance of Long Span Buildings Using Geometric Multi-Objective Optimization. Energy and Buildings, 127, 1–13.
20. Onatayo, D., Onososen, A., Oyediran, A. O., Oyediran, H., Arowoiya, V., & Onatayo, E. (2024). Generative AI Applications in Architecture, Engineering, and Construction: Trends, Implications for Practice, Education & Imperatives for Upskilling—A Review. Architecture, 4(4), 877-902